\documentclass[aps,preprint,onecolumn,11pt,showkeys]{revtex4}
\usepackage{amsfonts}
\usepackage{amsmath}
\usepackage{amssymb}
\usepackage{graphicx}
\usepackage[ansinew]{inputenc}
\usepackage[usenames,dvipsnames]{pstricks}
\usepackage{subfigure}
\usepackage{pdfpages}
\usepackage{epsfig}
\usepackage{pst-grad}
\usepackage{pst-plot}
\usepackage[colorlinks,hyperindex]{hyperref}

\hypersetup{
colorlinks,
citecolor=blue,
linkcolor=black,
urlcolor=black,
}
\DeclareMathOperator{\e}{e}

\begin{document}

\title{Refinements of the Weyl pure geometrical thick branes from
information-entropic measure}
\author{R. A. C. Correa}
\email{rafael.couceiro@ufabc.edu.br}
\affiliation{UNESP, Universidade Estadual Paulista, Campus de Guaratinguetá, 12516-410,
Guaratinguetá, SP, Brazil}
\author{D. M. Dantas}
\email{davi@fisica.ufc.br }
\affiliation{Universidade Federal do Ceará, 60455-760, Fortaleza, CE, Brazil}
\author{P.H.R.S. Moraes}
\email{moraes.phrs@gmail.com}
\affiliation{Instituto Tecnológico de Aeronáutica (ITA)- Departamento de Física,
12228-900, São José dos Campos, São Paulo, Brazil}
\author{A. de Souza Dutra}
\email{dutra@feg.unesp.br}
\affiliation{Universidade Estadual Paulista (UNESP), Campus de Guaratinguetá, 12516-410,
Guaratinguetá, SP, Brazil}
\author{C. A. S. Almeida}
\email{carlos@fisica.ufc.br }
\affiliation{Universidade Federal do Ceará (UFC), 60455-760, Fortaleza, CE, Brazil}
 
\begin{abstract}
This letter aims to analyse the so-called configurational entropy in the
Weyl pure geometrical thick brane model. The Weyl structure plays a
prominent role in the thickness of this model. We find a set of parameters
associated to the brane width where the configurational entropy exhibits
critical points. Furthermore, we show, by means of this
information-theoretical measure, that a stricter bound on the parameter of
Weyl pure geometrical brane model arises from the CE.
\end{abstract}

\maketitle

\section{INTRODUCTION}

Recently, the concept of entropy was reintroduced in the literature, by
taking into account the dynamical and the informational contents of models
with localized energy configurations \cite{gleiser-stamatopoulos}. Based on
the Shannon's information framework \cite{shannon}, which represents an
absolute limit on the best lossless compression of communication, the
Configurational Entropy (CE) was constructed. It can be applied to several
nonlinear scalar field models featuring solutions with spatially-localized
energy. As pointed out in \cite{gleiser-stamatopoulos}, the CE can resolve
situations where the energies of the configurations are degenerate. In this
case, it can be used to select the best configuration. The approach
presented in \cite{gleiser-stamatopoulos} has been used to study the
non-equilibrium dynamics of spontaneous symmetry breaking \cite%
{PRDgleiser-stamatopoulos}, to obtain the stability bound for compact
objects \cite{PLBgleiser-sowinski}, to study the bounds of Gauss-Bonnet
braneworld models \cite{rafael-tobias}, to explore dynamical holographic
AdS/QCD models \cite{roldao-alex}, to investigate the emergence of localized
objects during inflationary preheating \cite{PRDgleiser-graham} and to
distinguish configurations with energy-degenerate spatial profiles as well 
\cite{Rafael-Dutra-Gleiser}.

An interesting application of CE occurs in the context of braneworld models,
where the parameters of the scenarios can be bounded \cite{bc,
Rafael-Pedro,Rafael-Pedro2, Rafael-Davi}. In the braneworld perspective, the
Randall-Sundrum models \cite{RS1,RS2} assume that our observable universe is
a membrane embedded in an anti-de Sitter five-dimensional spacetime. From
this point of view, some prominent results were achieved as the resolution
of hierarchy problem \cite{RS1}, recovering of Newton's law \cite{RS2} and
Coulomb's law \cite{cw2, hung, Davi7} besides many applications to Cosmology
and High Energy Physics \cite{mc/2016,rscosmology, csaki1}. The recent
results in the diphoton experiment by ATLAS \cite{LHC1} and CMS \cite{LHC2}
brought a new phenomenological approach to these braneworld models \cite%
{750Gev, Han, radion, radion2, Csaki2016}. It has also been shown that the
braneworld might induce some observational effects in the gravitational wave
sign analysis \cite{mm/2014}.

In this work, we analyse the CE in the pure geometrical thick brane of
Reference \cite{Weyl2006}. A manifold endowed with Weyl structure is
considered, wherein Weyl scalar employs the thickness of the brane \cite%
{Weyl2006, W1,W2}. For this reason, the building of these Weyl thick brane
models \cite{ad1, ad2, ad3} arise naturally without the necessity of
introducing them \textquotedblleft by hand". This procedure preserves the 4D
Poincaré invariance and breaks $Z_{2}$-symmetry along the extra dimension.
The confinement of gravity in the Weyl brane was performed in \cite%
{Weyl2006,W1,W2}, while the matter fields in Reference \cite{Weyl2006b}.
Moreover, these pure geometrical thick branes remove the singularities and
solve the field non-confinement issues present in Randall-Sundrum thin
scenarios \cite{Bajc,5Dthick}.

In the next section of this work, we present a brief review of the Weyl
thick brane construction. The concept of CE and its application on the Weyl
thick brane are presented in Sec.\ref{s-2}. We finish with the results and
conclusions in Sec.\ref{s-3}.

\section{Review of pure geometrical Weyl thick brane model}

\label{s-1}

In this section, in order to put our work both in a more pedagogical and
comprehensive form in such a way that offers a coherent and credible guide
for reader, we will introduce a brief review about pure geometrical Weyl
thick brane. Thus, let us begin with the following Weyl (non-Riemannian)
action for a pure geometrical 5D scenario \cite{Weyl2006} 
\begin{equation}
S_{5}^{W}=\frac{1}{2\kappa }\int_{M_{5}^{W}}dx^{5}\sqrt{\lvert g\lvert }\e^{%
\frac{3\omega }{2}}\left[ R+3\tilde{\xi}\left( \nabla \omega \right)
^{2}+6U(\omega )\right] ,  \label{actionW}
\end{equation}%
where $g$ is the determinant of the metric $g_{MN}$ with $M,N$ running from $%
1$ to $5$ and $\omega $ represents the Weyl scalar function, which comprise
the pair for the Weyl manifold ($M_{5}^{W}$). The constant $\kappa =8\pi
G_{5}$ is related to the Newton constant in 5D, $R$ is the scalar curvature, 
$U(\omega )$ is the self-interacting potential of the scalar field and $%
\tilde{\xi}$ is an arbitrary parameter which enlarges the class of
potentials for which the 4D gravity can be localized. At this point, we
would like to emphasize that there are several physical motivations to study
the above theory. The first comes from the fact that geometrical thick
branes arise naturally without the necessity of introducing them by hand in
the action of the theory. In this case, spacetime structures with thick
smooth branes separated in the extra dimension arise, where the massless
graviton is located in one of the thick branes at the origin, meanwhile the
matter degrees of freedom are confined to the other brane. Another
remarkable motivation it was introduced by Drechsler \cite{refad}, where
such scenario can generate nonzero masses within the framework of a broken
gauge theory containing as a subsymmetry the electroweak gauge symmetry
which is known to contain many features in accord with observation.

Assuming for the line element in this 5D geometry an \emph{ansatz} of the
form 
\begin{equation}
ds^{2}=\e^{2A(y)}\eta _{\mu \nu }dx^{\mu }dx^{\nu }+dy^{2}\ ,
\label{5dmetric}
\end{equation}%
with $\eta _{\mu \nu }$ being the Minkowski metric with $\mu ,\nu $ running
from $1$ to $4$, $y$ the fifth dimension, and $A(y)$ the warp factor, the
stress-energy tensor components are obtained as \cite{Weyl2006,W1,W2} 
\begin{equation}
T_{mn}=\frac{3}{\kappa }\e^{2A}\left[ A^{\prime \prime }+2\left( A^{\prime
}\right) ^{2}\right] \eta _{mn},\quad T_{55}=\frac{6}{\kappa }\left(
A^{\prime }\right) ^{2}\,  \label{tmn}
\end{equation}%
with primes indicating derivatives with respect to the extra coordinate.

Now, mapping the Weylian action of Eq.\eqref{actionW} into the Riemannian
one through the conformal transformation $\hat{g}_{MN}=\e^{\omega }\eta _{MN}
$, we have \cite{Weyl2006,W1,W2} 
\begin{equation}
S_{5}^{R}=\frac{1}{2\kappa }\int_{M_{5}^{W}}dx^{5}\sqrt{\lvert g\lvert }%
\left[ R+3\xi \left( \hat{\nabla}\omega \right) ^{2}+6\hat{U}(\omega )\right]
\ ,  \label{actionR}
\end{equation}%
where the transformation changes the terms $\xi =\tilde{\xi}-1$ and $\hat{U}%
(\omega )=\lambda \e^{\frac{4k\xi \omega }{1+k}}U(\omega )$. The metric of
Eq.\eqref{5dmetric} becomes 
\begin{equation}
ds^{2}=\e^{\omega (y)}\left[ \e^{2A(y)}\eta _{\mu \nu }dx^{\mu }dx^{\nu
}+dy^{2}\right] \   \label{metric}
\end{equation}%
in this Riemannian frame.

In order to work with a first order differential system, the authors of Ref.%
\cite{Weyl2006, W1,W2} write $X=\omega ^{\prime }$ and $Y=2A^{\prime }$.
Thus, we have the following pair of coupled field equations\textbf{\ }%
\begin{eqnarray}
X^{\prime }+2YX+\frac{3}{2}X^{2} &=&\frac{1}{\xi }\frac{d\hat{U}}{d\omega }%
e^{\omega }, \\
Y^{\prime }+2Y^{2}+\frac{3}{2}XY &=&\left( -\frac{1}{\xi }\frac{d\hat{U}}{%
d\omega }+4\hat{U}\right) e^{\omega }.
\end{eqnarray}

Note that the above equations are nonlinear, and unfortunately, as a
consequence of the nonlinearity, in general we lose the capability of
getting the complete solutions. However, it was shown in Ref. \cite{W1} that
a special class of analytical solution can be easily obtained with the
imposition $X=kY$, where $k\neq 1$ is an arbitrary constant parameter. Thus,
both above equations are reduced to the following single differential
equation 
\begin{equation}
Y^{\prime }+\frac{4+3k}{2}Y^{2}=\frac{4\lambda }{1+k}\e^{\omega \left( \frac{%
4k\xi }{1+k}+1\right) }\ .  \label{edo}
\end{equation}

Looking for an alternative solution of Ref.\cite{W1,W2}, where $\xi =\frac{%
1-k}{4k}$ and $U=\lambda $ (being $\xi $ determined by arbitraries $k$), the
Ref.\cite{Weyl2006} sets $\xi =-7/16$, which leads to a potential of the
form $U(\omega )=\lambda \e^{-6\omega }$. This potential breaks the
invariance under Weyl scaling transformations for arbitrary $\xi \neq 0$ and
also transforms the geometrical scalar field into an observable one \cite%
{Weyl2006}. From the impositions of Ref.\cite{Weyl2006} the differential
equation \eqref{edo} yields to 
\begin{equation}
Y^{\prime }=-12\lambda \e^{-p\omega }\quad \text{or}\quad \omega ^{\prime
\prime }=16\lambda \e^{-p\omega },  \label{edo2}
\end{equation}%
where $p=1+16\xi $, which leads to the relation $A^{\prime }=\frac{3}{8}%
\omega ^{\prime }$.

Hence the functions $A(y)$ and $\omega (y)$ are determined as 
\begin{equation}
\e^{2A(y)}=\left\{ \frac{\sqrt{-8\lambda p}}{c_{1}}\cosh \left[
c_{1}(y-c_{2})\right] \right\} ^{\frac{3}{2p}}\ ,\quad \omega (y)=-\frac{2}{p%
}\ln \left\{ \frac{\sqrt{-8\lambda p}}{c_{1}}\cosh \left[ c_{1}(y-c_{2})%
\right] \right\} \ ,  \label{ay}
\end{equation}%
where $c_{1}$ (related to the inverse of warp-factor width) and $c_{2}$
(which is associated to where the warp-factor centering located) are
arbitrary integration constants.

Thus, with the $A(y)$ form above, the energy density obtained from $\rho
(y)=T_{00}(y)$ is given by 
\begin{equation}
\rho (y)=-\frac{9c_{1}^{2}}{4\kappa p}\left\{ \frac{\sqrt{-8\lambda p}}{c_{1}%
}\cosh \left[ c_{1}(y-c_{2})\right] \right\} ^{\frac{3}{2p}}\left\{ 1+\frac{%
3-2p}{2p}\tanh ^{2}\left[ c_{1}(y-c_{2})\right] \right\} .  \label{energy}
\end{equation}

For the non-compact extra dimension case (Randall-Sundrum 2 model \cite{RS2}%
) where $\lambda>0$, $c_1>0$ and $p<0$, this quantity will be localized and
centered at $c_2$, while $c_1$ regulates the maximum amplitude of this
energy. The thin Randall-Sundrum 2 limit for this distribution is obtained
when $p\to-\infty$ and $c_1\to\infty$ (provided that $-p/c_1$ is finite). It
is important to comment that for this scenario the usual $Z_2$ symmetry
along the $y$ coordinate is not necessarily preserved \cite{Weyl2006}. We
plot the warp-factor of Eq.\eqref{ay} and the energy density of Eq.%
\eqref{energy} in Fig.\ref{fig-warp-energy} below.


\begin{figure}[!htb]
\includegraphics[width=.48\linewidth]{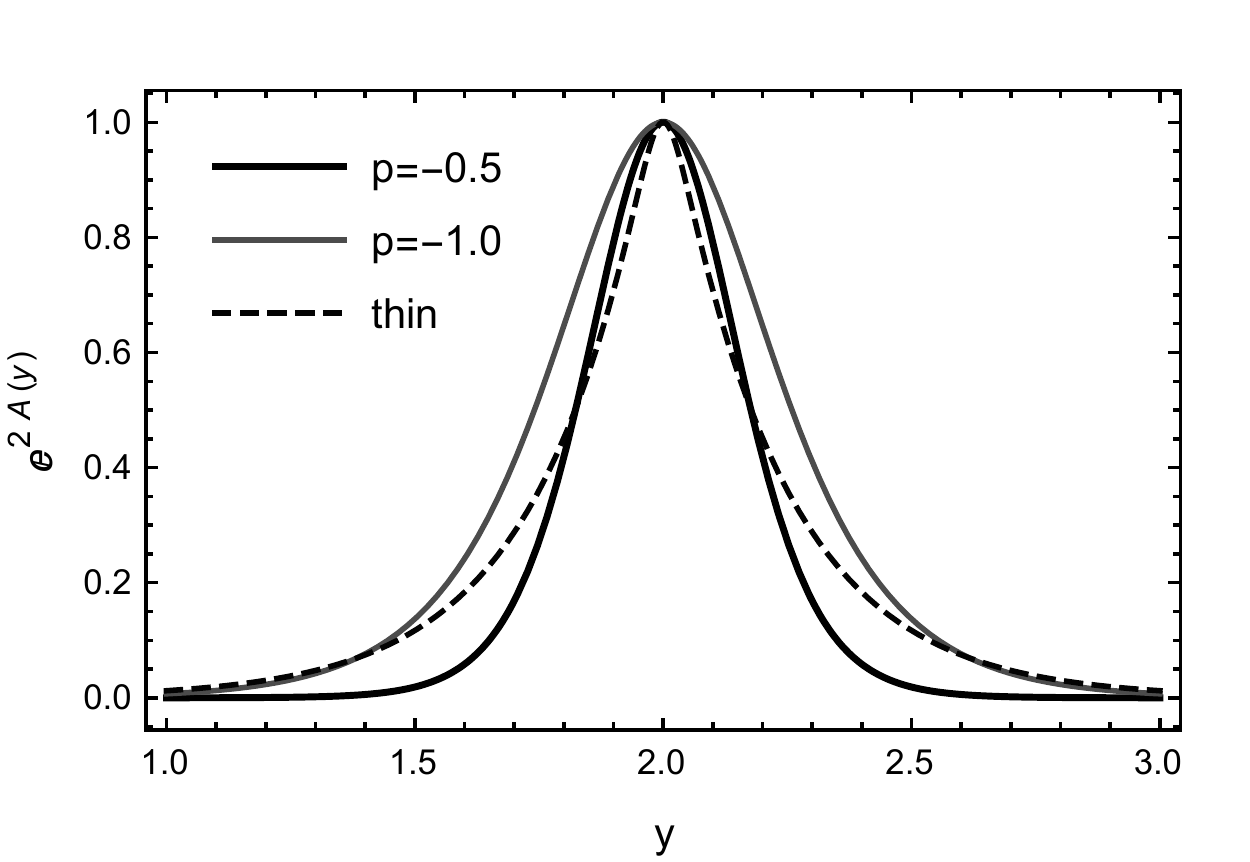} %
\includegraphics[width=.48\linewidth]{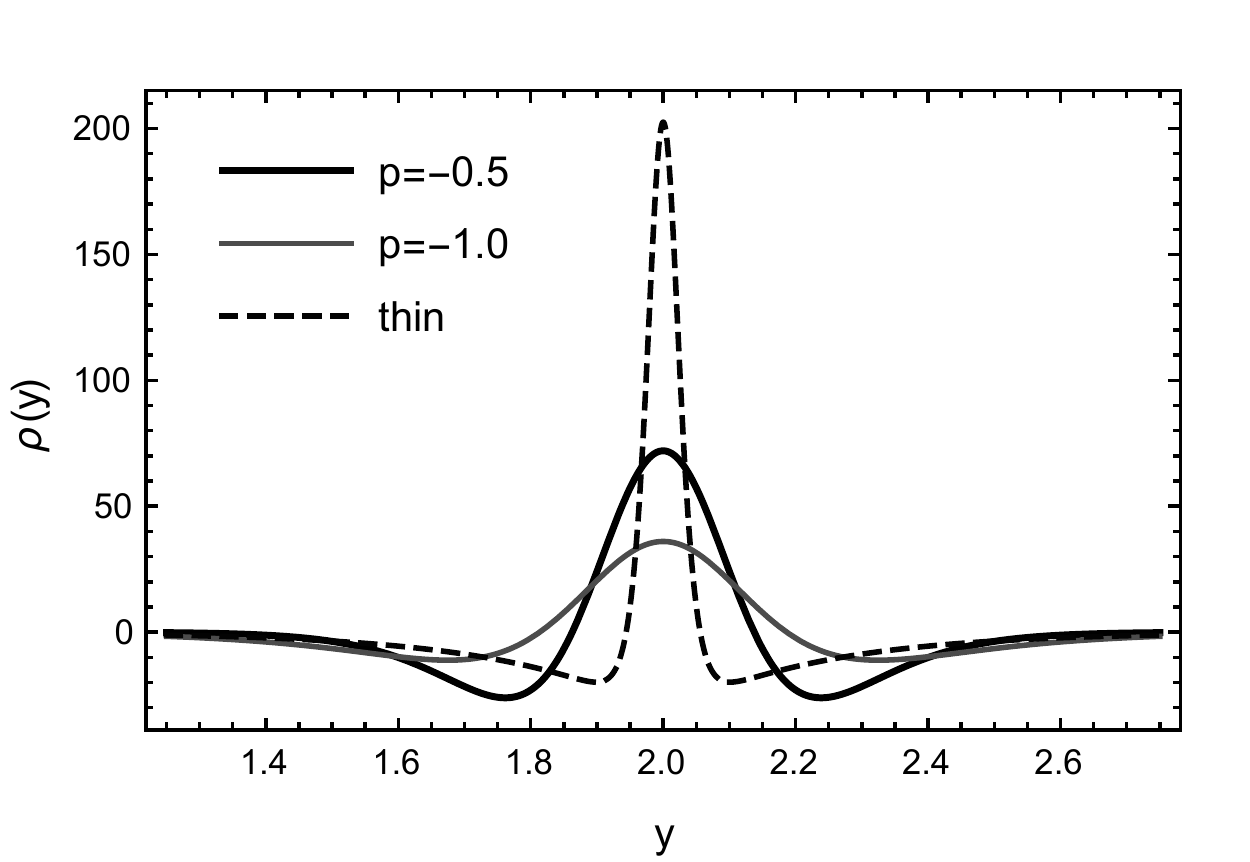}
\caption{Warp-factor (left) and energy density (right) for the non-compact
case. We made $\protect\kappa=1$, $\protect\lambda=-\frac{c_1^2}{8p}$ and $%
c_2=2$ (which centralized the brane around $y=2$). The thick branes are
displayed with $c_1=4$ with $p=-0.5$ (thick black line) and $p=-1.0$ (thin
grey line). For qualitative thin brane (dashed line) we set $p=-\frac{c_1}{3}%
=-10$.}
\label{fig-warp-energy}
\end{figure}


At this point, it is important to remark that despite having a set of
analytical solution, a complete description of the question of bounds on the
parameters of the solution, remains an open issue. Then, in the next
sections, we seek to address this gap in the literature by finding through
of configurational entropy concept a stricter bound on the parameter of
these solutions and its physical relevance.

\section{Configurational Entropy approach}

\label{s-2}

Gleiser and Stamatopoulos \cite{gleiser-stamatopoulos} have recently
proposed a detailed picture of the CE for the structure of localized
solutions in classical field theories. In this section, analogously to that
work, we formulate a CE measure in the functional space, from the field
configurations where the pure geometrical Weyl thick brane scenarios can be
studied. Firstly, the framework shall be formally introduced and thereafter
its consequences will to be explored.

{{There is an intimate link between information and dynamics, where the
entropic measure plays a prominent role. The entropic measure is well known
to quantify the informational content of physical solutions to the equations
of motion and their approximations, namely, the CE in functional space \cite%
{gleiser-stamatopoulos}. Gleiser and Stamatopoulos proposed that nature
optimises not solely by optimising energy through the plethora of \emph{a
priori} available paths, but also from an informational perspective \cite%
{gleiser-stamatopoulos}. }}

To start, let us write the following Fourier transform 
\begin{equation}
\mathcal{F}[\Omega]=-\frac{1}{\sqrt{2\pi }}\int dr\;e^{i\Omega r}\mathcal{L},
\label{3.1}
\end{equation}
where $\mathcal{L}$ is the standard Lagrangian density.

Now the modal fraction is defined by the following expression \cite%
{gleiser-stamatopoulos, PLBgleiser-sowinski, Rafael-Dutra-Gleiser}:%
\begin{equation}
f(\Omega )=\frac{\left\vert \mathcal{F}[\Omega ]\right\vert ^{2}}{\int
d\Omega \left\vert \mathcal{F}[\Omega ]\right\vert ^{2}}  \label{3.2}
\end{equation}%
and measures the relative weight of each mode $\Omega $.

{The CE was motivated by the Shannon's information theory. Indeed, it was
originally defined by the expression $S_C[f] = -\sum f_n\,\ln(f_n)$, which
represents an absolute limit on the best lossless compression of any
communication \cite{shannon}. The CE thus originally provided the
informational content of configurations that are compatible to constraints
of an arbitrary physical system. When $N$ modes $k$ carry the same weight,
then $f_n = 1/N$ and hence the discrete CE has a maximum at $S_C = \ln N$.
Instead, if just one mode constitutes the system, then $S_C = 0$ \cite%
{gleiser-stamatopoulos}.}


{Analogously, for arbitrary non-periodic functions in an open interval, the
continuous CE can be described} by the expression%
\begin{equation}
S_{c}[f]=-\int d\Omega f(\Omega )\ln [f(\Omega )].  \label{3.3}
\end{equation}


Thus, Eq.(\ref{3.1}) can be used to generate the normalized fraction, in
order to obtain the entropic profile of thick brane solutions. Furthermore,
we must find $F(\vec{k})$, defined by the Fourier transform 
\begin{equation}
F(\vec{\Omega})=\frac{1}{(2\pi )^{d/2}}\int d^{d}x\;e^{i\vec{\Omega}\cdot {%
\vec{x}}}\,\rho (\vec{x})\,.\qquad  \label{54}
\end{equation}%
\noindent As we are dealing with one spatial dimension, the above expression
reads: 
\begin{equation}
F({\Omega })=\frac{1}{\sqrt{2\pi }}\int dx\;e^{i\Omega y}\,\rho ({x})\,.
\end{equation}

In order to completely specify the normalized fraction $f(\Omega )$, we must
calculate $\int \;dk|F(\Omega )|^{2}.$ Hence, by using the Plancherel
theorem it follows that%
\begin{equation}
\int dk\left\vert {F}(\Omega )\right\vert ^{2}=\int dx\left\vert \rho
(x)\right\vert ^{2}.  \label{plancherel}
\end{equation}

Thus, in the next section, we will apply this new approach to investigate
pure geometrical Weyl thick brane theories, which is given by Eq. (\ref%
{actionW}). As we will see, important consequences will arise from the CE
concept. Furthermore, we will show that the CE provides a stricter bound on
the parameters of the related partial diferential equation (PDE) solution
which is represented in Eq. (\ref{ay}).

\section{Configurational Entropy in the Weyl Brane}

\label{s-3.1}

We now use the approach presented in the previous section to obtain the CE
of the Weyl brane configurations, which has the set of exact solutions,
given in Section II. Let us begin by rewriting the energy density of Eq.%
\eqref{energy} in the form 
\begin{equation}
\rho (y)=A_{1}\left\{ \cosh [c_{1}(y-c_{2})]\right\} ^{a}+A_{2}\left\{ \cosh
[c_{1}(y-c_{2})]\right\} ^{b}\left\{ \sinh [c_{1}(y-c_{2})]\right\} ^{2},
\label{1}
\end{equation}%
where we are using the following definitions%
\begin{equation}
A_{1}\equiv -\frac{9c_{1}^{2}}{4p}\kappa \left( \frac{\sqrt{-8\lambda p}}{%
c_{1}}\right) ^{a},\quad \ A_{2}\equiv A_{1}\left( \frac{6-2p}{2p}\right) ,
\end{equation}%
with $a=3/2p$ and $b=3/2p-2$. Thus, we have%
\begin{equation}
\mathcal{F}(\Omega )=\int dy\rho (y)e^{i\Omega y}.
\end{equation}

In this case, we obtain%
\begin{equation}
\mathcal{F}(\Omega)=\frac{e^{i\Omega c_{2}}}{c_{1}}\sum\limits_{m=1}^{2}A_{m}%
\mathcal{F}_{m}(\Omega),  \label{2}
\end{equation}

where%
\begin{eqnarray}
&&\left. \mathcal{F}_{1}(\Omega)=\frac{2^{-a-3/2}i^{a}}{\sqrt{\pi }}%
\sum\limits_{j=1}^{2}e^{(-1)^{j}\Omega\pi /c_{1}}\beta _{-1}\left[ -\frac{a}{%
2}+\frac{(-1)^{j}\Omega}{2c_{1}},1+a\right] ,\right. \\
&&  \notag \\
&&\left. \mathcal{F}_{2}(\Omega)=\frac{2^{-a-5/2}}{\sqrt{\pi }}\left\{
2^{b+2}\sum\limits_{j=1}^{2}\frac{1}{b+(-1)^{j+1}ik/c_{1}}H\left[ 1,1+\frac{b%
}{2}+\frac{(-1)^{j}i\Omega}{2c_{1}};1-\frac{b}{2}+\frac{(-1)^{j}i\Omega}{%
2c_{1}};-1\right] \right. \right.  \notag \\
&&  \notag \\
&&\left. -\sum\limits_{j=1}^{2}\frac{1}{2+b+(-1)^{j+1}i\Omega/c_{1}}H\left[
-b,-1-\frac{b}{2}+\frac{(-1)^{j}i\Omega}{2c_{1}};-\frac{b}{2}+\frac{%
(-1)^{j}i\Omega}{2c_{1}};-1\right] \right.  \notag \\
&&  \notag \\
&&\left. \left. +\sum\limits_{j=1}^{2}\frac{1}{2-b+(-1)^{j+1}i\Omega/c_{1}}H%
\left[ -b,1-\frac{b}{2}+\frac{(-1)^{j}i\Omega}{2c_{1}};2-\frac{b}{2}+\frac{%
(-1)^{j}i\Omega}{2c_{1}};-1\right] \right\} .\right.
\end{eqnarray}

In the above equations, $H[M_{1},M_{2};M_{3;}M_{4}]$ are the hypergeometric
functions and $\beta _{-1}(X,Y)$ is the incomplete beta function. 

\begin{figure}[!htb]
\begin{minipage}[t]{0.45 \linewidth}
\includegraphics[width=1.\linewidth]{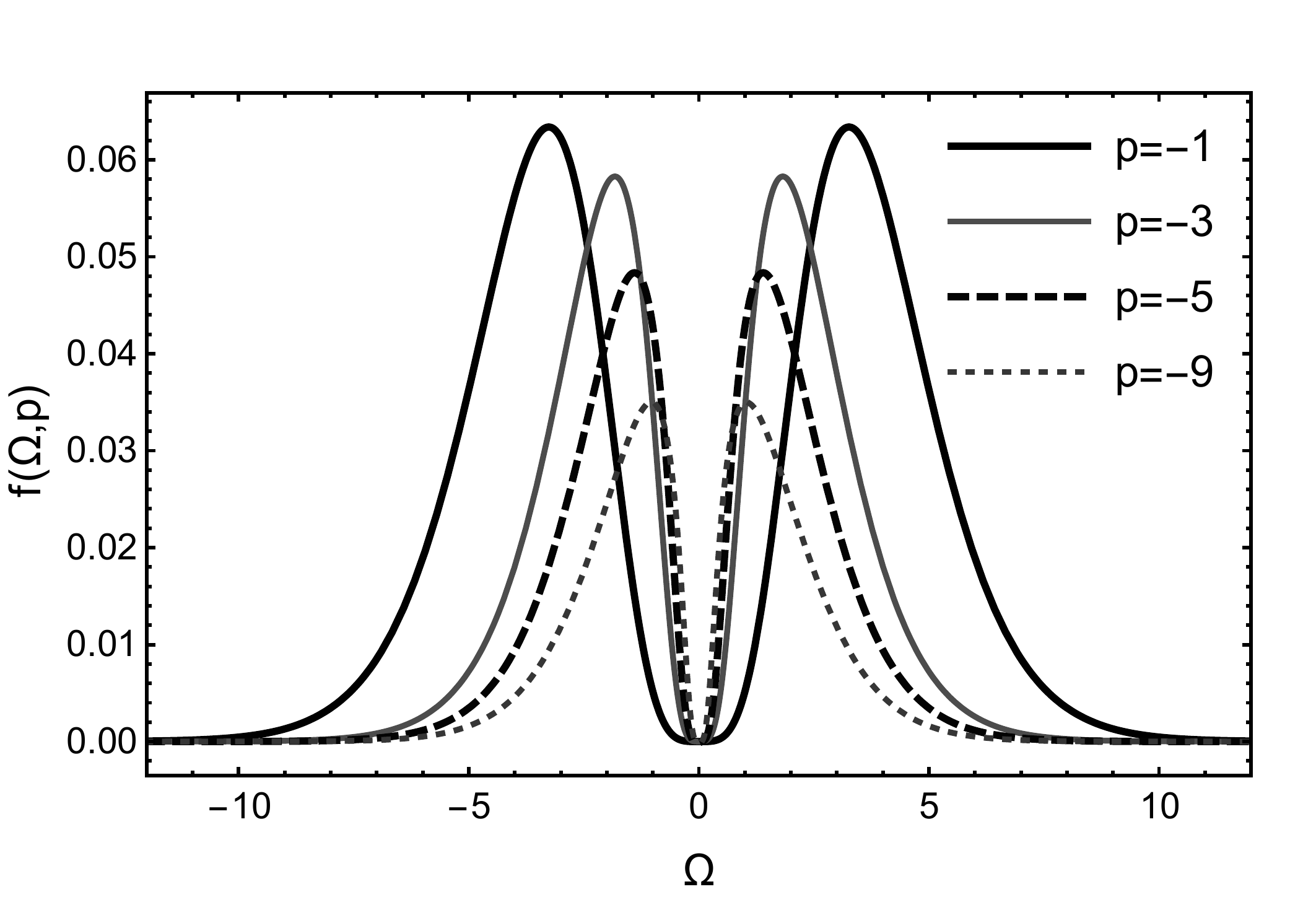}
\caption{$f(\Omega)$: normalized fraction for the Weyl thick brane as function of associated coupling parameter $p=1+16\xi $. The parameter $p$ regulates the amplitude of $f(\Omega)$.} 
\label{fig-f-p}
\end{minipage}
\quad 
\begin{minipage}[t]{0.45 \linewidth}
\includegraphics[width=1.\linewidth]{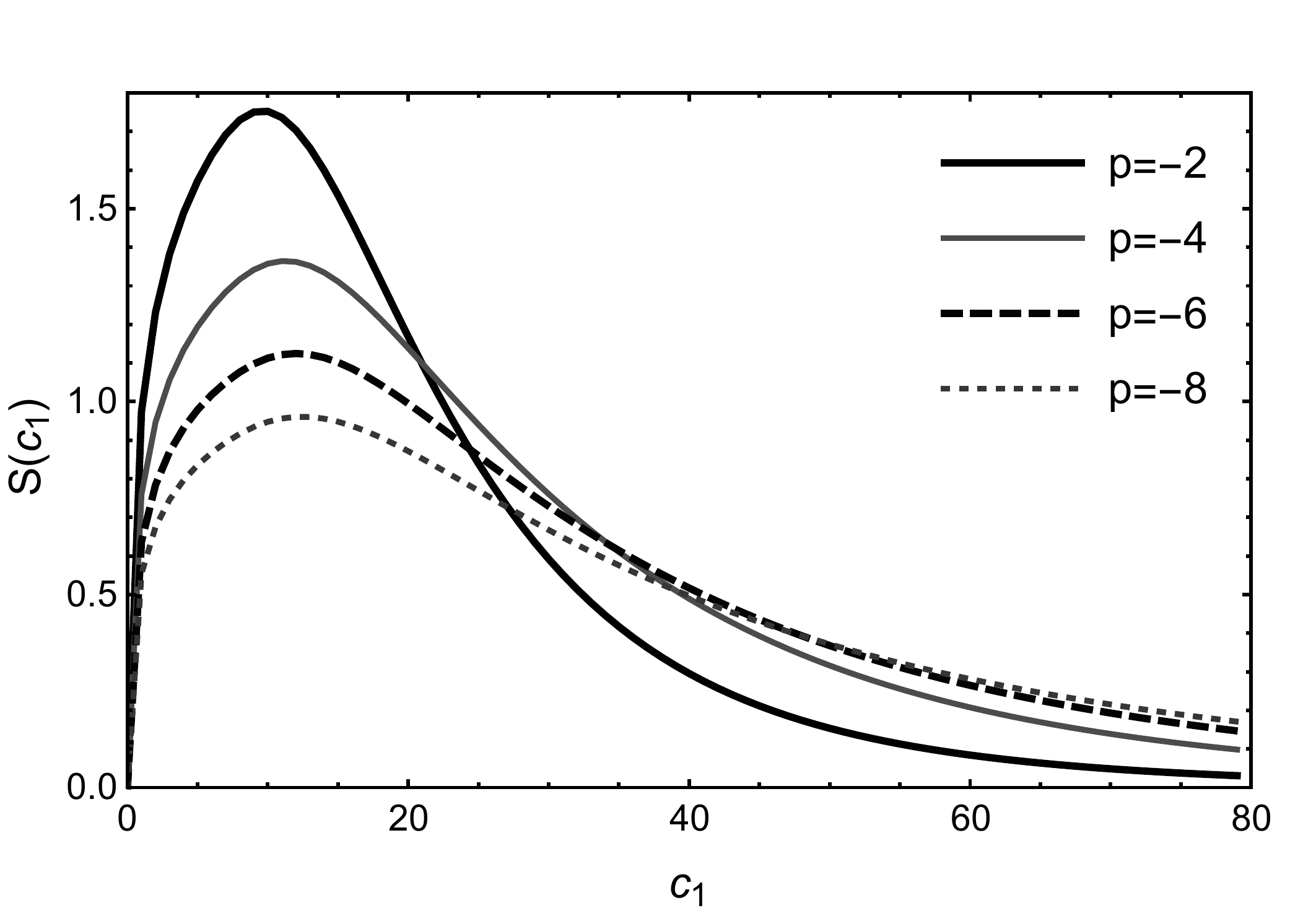}
\caption{$S(c_1)$: CE for the Weyl thick brane. When $p\to 0$ and the thin case (where $p\to - \infty$) the CE tends to zero. The CE has a critical maximum detailed in Fig.\ref{fig-c1-crit}.} 
\label{fig-s-p} 
\end{minipage}                
\end{figure}
\begin{figure}[tbh]
\includegraphics[width=.5\linewidth]{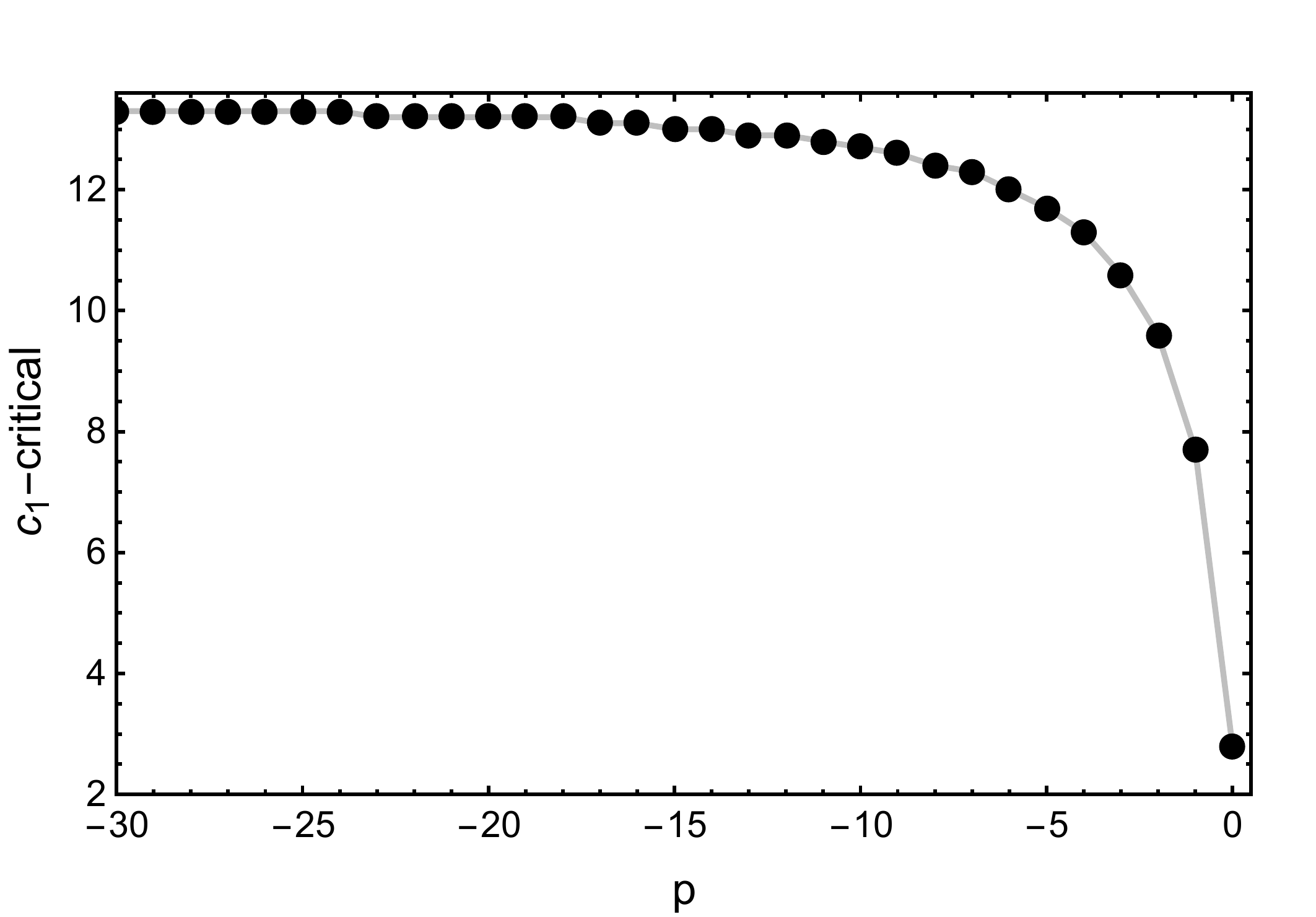}
\caption{Critical point of CE: when $p\rightarrow -\infty $, $%
c_{1}\rightarrow 13$.}
\label{fig-c1-crit}
\end{figure}

As a result, we plot the normalized fraction in Fig. \ref{fig-f-p} where we
verify that the Weyl parameter $p$ is inversely proportional to the
amplitude of $f(\Omega )$. The CE is represented in Fig. \ref{fig-s-p},
where we note that the Weyl parameter decreases the amplitude of the CE.
There is a critical point of the CE as a function of the thickness parameter 
$c_{1}$, as also verified in other thick braneworld models \cite{bc,
Rafael-Pedro,Rafael-Pedro2, Rafael-Davi}. The evolution of this critical
point is represented in Fig.\ref{fig-c1-crit}.


As we can see, considering the 5-dimensional Weyl gravity model which is
given by Eq. (\ref{actionW}), it was possible to obtain the
information-entropy measure for the system under analysis. Here, it is
important to remark that the CE concept states that there is an intimate
link between information and dynamics, where entropic measure is responsable
in quantify the informational content of physical solutions to the equations
of motion and their approximations. In this case, from Fig. \ref{fig-s-p} we
can verify that there is a sharp minimum at the value $c_{1}\rightarrow
\infty $, the region of parameter space where the brane solutions are most
prominent from the informational point of view. Therefore, in this scenario,
the information entropy-measure provide that, when the warp factor of the
brane is concentrated near $y=c_{2}$, the corresponding warp factor
reproduces the metric of the Randall-Sundrum model in the thin brane limit,
showing that the scalar curvature of the Weyl integrable manifold turns out
to be completely regular in the extra dimension. Furthermore, at large
values of $c_{1}$, the brane-world CE yields the configurational entropy $%
S_{c}(c_{1})\rightarrow 0$, showing a great organisational degree in the
structure of configuration of the system. Finally, by using the approach
presented by GS \cite{gleiser-stamatopoulos}, we have checked that the CE of
the configurations given in Section II, is correlated to the energy of the
system. The higher [lower] the brane configurational entropy, the higher
[lower] the energy of the solutions.

\section{DISCUSSION and CONCLUSIONS}

\label{s-3}

In this work we performed the analyses of the CE in the Weyl pure
geometrical thick brane. This braneworld model was presented in Refs.\cite%
{Weyl2006,W1,W2}, where the thickness of the model arise from the Weyl
scalar fields. These models are subject to some parameter, in special those
that regulates the model width as the $c_1$ (an arbitrary integration
constant) and $p$ (another arbitrary constant associated which is associated
to the Weyl coupling parameter $\xi$ \cite{Weyl2006,W1,W2}).

Applying the CE concept in this model, we conclude that for a fixed $p$ the
CE produces a minimum when $c_{1}\rightarrow 0$ (large thickness) and $%
c_{1}\rightarrow \infty $ (thin limit). On the other hand, a maximum point
in the CE is observed for a specific arrangement of the pair $c_{1}$ and $p$%
. Here, it is important to highlight that our results are consistent with
those shown in \cite{Rafael-Pedro,Rafael-Davi,rrd}.

One important physical conclusion that we can draw from our study of the
Weyl brane structures, which is given for the model presented in Section II,
is the fact that the CE approach can be used to split the solution in two
classes. The first one, is given at the values $0<c_{\mathbf{1}}\lesssim 100$%
, where the configurational entropy measure has values higher than $0$. In
this case, the configurational entropy suggest that the Weylian scalar
curvature is non--singular along the fifth dimension. On the other hand,
when $c_{1}\rightarrow \infty $, the configurational entropy indicates that $%
S_{c}\rightarrow 0$, and the 5--dimensional curvature scalar is singular. As
a consequence, the Randall-Sundrum model is recovered.

Another remarkable result in our analysis comes from the thickness of the
domain wall, which is given by the parameter $p$. It is straightforward to
realise from Fig. \ref{fig-s-p} that the higher the brane thickness $p$, the
higher the respective value for the brane configurational entropy as well.
Moreover, the Weyl brane-world configurational entropy is used to evince a
higher organisational degree in the structure of system configuration
likewise, for large values of $c_{1}$.

We also intend to study the CE in the six dimensional scenarios of Refs.\cite%
{D1, D2}, in addition to those studied in Refs. \cite{GS1, Torrealba, D3, D4}%
, presented in the recent work \cite{Rafael-Davi}.\newline

\textbf{Acknowledgments}

RACC, DMD and CASA thank Coordenação de Aperfeiçoamento de Pessoal de Nível
Superior (CAPES), Conselho Nacional de Desenvolvimento Científico e Tecnoló%
gico (CNPq) and Fundação Cearense de apoio ao Desenvolvimento Científico e
Tecnológico (FUNCAP) for financial support. RACC also acknowledges
Universidade Federal do Ceará (UFC) for the hospitality. PHRSM would like to
thank S\~ao Paulo Research Foundation (FAPESP), grant 2015/08476-0, for
financial support. 

\end{document}